\documentclass[11pt]{article}
\usepackage[a4paper,margin=2.5cm]{geometry}
\usepackage[T1]{fontenc}
\usepackage{lmodern}
\usepackage{amsmath,amssymb}
\usepackage{booktabs,array,tabularx}
\usepackage[round,semicolon,authoryear]{natbib}
\bibpunct{(}{)}{;}{a}{,}{,}
\usepackage[labelfont=bf,figurename=Fig.,labelsep=space]{caption}
\usepackage{graphicx}
\usepackage{enumitem}
\usepackage{hyperref}
\hypersetup{citebordercolor=0 1 0,linkbordercolor=0 1 0,urlbordercolor=0 1 0}
\title{Which Policy Effects Do Causal Estimators Recover?\\Evidence from an Agent-Based Cournot Market with Learning Firms}
\author{Beril Demiroglu\\Department of Economics, Y\i ld\i z Technical University}
\date{}

\begin{document}
\maketitle

\begin{abstract}
The impact on the whole market of the policy that changes the competitive behavior cannot be obtained from the comparison of the treated firms and the untreated firms. In our study, we perform the simulations for the agent-based market where agents play a Cournot game and use Q-learning. Here, the impact of the treatment on the firm affects the market price and, consequently, impacts the untreated firms. Next, using our counterfactual framework, we estimate four different impacts: direct impact, total impact of the treatment on the treated firms, spillover impact of the treatment on the untreated firms, and total impact of the treatment for everyone. Finally, we assess which of these effects could be estimated by methods such as naive contrast, within-market regression, difference-in-differences, double machine learning, causal forest, and exposure-aware versions of the last two. In the markets with 4 to 8 firms, slow learning, and no selection into treatment, the within-market estimators measure the total effect of the treatment for the treated firms, minus the spillover effect, and the average absolute bias against the direct effect is from 12\% to 22\%. The global effect could not be estimated when the treated proportion is the same in all the markets, but when it is not, the exposure-aware estimators estimate it within 10 percent without selection into treatment, and with selection the bias grows with the market size. Double machine learning and causal forest produce similar results. The simulation design, code, and protocol are available along with the paper.


\end{abstract}

\section{Introduction}
If a policy changes one unit's cost structure, that will affect its output level and the market price. Therefore, every firm faces the price change, which means that the effect of the subsidy on the treated firm depends on the number of rival firms treated and untreated firms are affected as well. The basic comparison of the treated and untreated entities fails to capture the effect of the policy on the whole market because of an issue called interference. There is substantial literature on the topic \citep[see][]{hudgens2008,aronow2017,baird2018,savje2021,hu2022}. For the markets where there is interference via prices, \citet{mkw2025} show that the direct effect and the effect of treating all entities can be far apart and provide estimators and designs for large markets consisting of price-takers.

Here we ask an additional question that these findings leave open -- in a small market with firms that have market power and learn, which possible effects can be captured by standard estimators? Real-world data is unable to answer this because the effect of treating everyone is never observed. Hence we construct a simulator where the true effects are known. Firms in a Cournot game learn their level of output via Q-learning; a per-unit subsidy is given to some firms; and the effects are computed in paired counterfactual worlds that share the same random draws. Then the estimators are applied to the data a researcher would observe (one treatment assignment per market) and compared with each true effect.

\textbf{Contributions.} (i) A map of which effect each estimator recovers, compared with the known truth, depending on the market size, the share of treated firms, and the selection into treatment. (ii) A measure of the gap between the within-market contrast and the direct effect, and of how this gap changes with the market size. (iii) Evidence on how much varying the treated share across the markets, which is an old idea in the design literature, helps the general-purpose estimators when firms learn. (iv) A simulator that can be used again, a placebo check, and an analysis protocol that we wrote down and committed before the main run.

\textbf{What this paper does not claim.} It is already known that the treated--control contrast differs from the global effect under interference, that this can happen through prices, and that the global effect can be estimated by randomizing the treated share across separate markets \citep{baird2018,mkw2025}. The contribution is quantitative and applies to the following setting: small Cournot markets with learning firms and general estimators. The results apply to the above setting (one calibration, stateless learners), and we do not make any claims about their applicability to real markets. Our literature review was not exhaustive, and therefore we do not make any claims about the existence of similar studies.

\section{Related work}
\textbf{Interference and designs.} \citet{hudgens2008} define direct, indirect, total and overall effects under interference within groups, and \citet{aronow2017} give a design-based framework with exposure mappings. \citet{baird2018} examine randomized saturation designs, where the proportion assigned to treatment differs across clusters. \citet{savje2021} distinguish between two definitions of an average direct effect under unit-interference: the effect of switching a unit's assignment, while the assignments of all other units remain constant, and the gap between the means of outcomes for those who were treated and those who were not, implying that the treatment and control groups face different distributions of assignments among other units. \citet{hu2022} draw a similar parallel with regard to the direct effect of \citet{hudgens2008}: it is equal to the average direct effect only in Bernoulli designs, and in complete randomization designs, it may be non-zero, despite no unit having its direct effect. \citet{leung2020} analyzes treatment and spillover effects under interference through a network, as well as dependence of a unit's outcome on the number of treated neighbors and the total number of neighbors; he develops the asymptotic results for large, sparse networks.
In his analysis of nonparametric exposure contrasts, \citet{leung2026} defines these as the contrasts of the mean outcomes between different exposure groups of units. He provides an example illustrating that in this case, even when assignment is unconfounded, the exposure contrast can have the opposite sign of every unit-level effect if the exposure mapping does not account for all possible interference and there exists correlation in assignment across units, and he presents conditions for either the interference or the assignment under which the contrast is a convex combination of unit-level effects. In his analysis, \citet{savje2024} distinguishes the two roles played by an exposure mapping: definition of the target effect and assumption that the mapping describes all interference. In his expected exposure effect, the researcher compares mean outcomes of units with different exposures and thus this measure depends on the design since conditioning on the exposure of one unit modifies the distribution of treatment of other units; the Horvitz--Thompson estimators are consistent for this effect provided positive probability of exposures and only weak dependence of the misspecification errors across units. He leaves the limiting distribution and variance estimation under misspecification open. \citet{manski2013} analyzes what potential outcomes distributions could be known if the outcomes of an individual would be affected by the treatments of the entire population and finds only bounds and not point estimates. He supposes that the outcomes are invariant across treatment assignments giving the individual the same effective treatment, for example, the same distribution of treatments within the individual's reference group, and finds out that the point identification of the distribution of outcomes under a specific treatment assignment involves every possible effective treatment in it to have some positive probability under the realized treatment assignment. For the treatment interactions extending to the entire population or to a distribution of treatments within the group, he mentions that the random assignment does not necessarily meet this requirement because the realized distribution of treatments is unlikely to coincide with the specific one. We read this as an instance of his argument: a fixed share of treated firms in each market cannot meet it for the all-treated case; he does not state this for markets. His analysis is on identification rather than estimation, and his model does not include price-mediated markets and learning agents. In this case, the market is the cluster while the proportion of treated competitors is the exposure.

\textbf{Prices as interference mechanisms.} \citet{mkw2025} study a randomized experiment in one big market where spillovers occur via observed prices. They define the average direct effect, the average indirect effect, a marginal policy effect, and the total effect. They show that a difference in means is consistent for the direct effect, that the total effect needs variation across disconnected markets, and that the indirect effect can be estimated by randomizing small price perturbations. Their theory is a large-market limit where the market power of a single unit disappears, and they note that firms with significant market power need different techniques. \citet{munro2026} studies designed markets with cutoff mechanisms, such as auctions and school choice. The paper reports that a forest-based doubly robust average-effect estimator, which ignores equilibrium effects, stays biased when the market grows. \citet{arkhangelsky2025} show that the effect of a marginal reform in a centralized market, such as an admission cutoff or a market-clearing price, can be identified from the data of one baseline market, but only as a local effect. They focus on identification, not estimation, and they leave out peer effects and strategic behavior. \citet{zhan2026} study creator-side experiments on a content platform where treated and control items compete for exposure. They show that difference-in-means can be severely biased, and they propose a structured model of the allocation mechanism together with a debiased (double machine learning) estimator of the global effect of rolling out to the whole platform. These papers propose estimators for settings where the interference channel is known; in this paper we compare general-purpose estimators when this channel is not imposed.

\textbf{Estimators.} We use the double machine learning with cross-fitting \citep{chernozhukov2018}, the causal forests \citep{wager2018,athey2019}, and difference-in-differences. \citet{knaus2021} compare such machine learning estimators in an empirical Monte Carlo study. In their study there is no interference and they look at point estimates, not inference. \citet{butts2023} studies difference-in-differences with spatially local spillovers. He shows that, under parallel counterfactual trends, the difference-in-differences estimand is the total effect on the treated minus the spillover on the controls. In a Cournot market every firm is a neighbor through the price, so no firm is far from the treatment. \citet{shirani2025} release six simulated environments with known truth for network interference, and one of them is a price-driven auction. They compare a causal message-passing estimator with difference-in-means and Horvitz--Thompson estimators. They do not evaluate double machine learning or causal forests, and none of their environments is a Cournot market with learning firms.

\textbf{Learning firms.} \citet{calvano2020} let Q-learning pricing algorithms with a one-period memory play a repeated price game with logit demand and find prices above the competitive level that are sustained by punishment of deviations. In addition, they point out that in case the algorithms are either myopic or lack memory, the only possible equilibrium of the repeated game is competitive and the algorithms learn to price near it. In their study of firms using Q-learning in the context of a repeated Cournot game with linear demand, \citet{waltman2008} find that the firms are producing lower amounts and earning higher profits compared to what is possible in the one-shot Nash equilibrium; the authors define any solution that earns firms joint profit greater than Nash equilibrium as collusive and they note that most of their firms cannot remember the past and therefore cannot punish. They have shown that for two firms having two levels of production and lack of memory, the firms spend almost all the time producing at the lower level of production as the exploration parameter in the logit choice rule approaches zero, which they attribute to the fact that exploration decreases when the difference between Q-values is high; in the case of two actions, they state that without proving their claim that simulations conducted using a fixed exploration probability tend to end up in the Nash equilibrium. According to \citet{calvano2020}, the above results without memory might be due to a lack of learning process and not collusion, and the interpretation is according to Calvano et al., not Waltman \& Kaymak, and \citet{klein2021} states the result similarly. Klein examines price setting firms operating in turn within a duopoly and finds out that Q-learning algorithms lead to supra-competitive prices or price cycles; in addition, he states that there is no convergence guarantee in the case of several Q-learners learning together. The learners here have no state and are characterized by $\gamma=0$ and therefore unable to represent punishment strategies, so the results obtained below are not collusion sustained by punishment; the players are only used as means for providing the necessary feedback loop in the market. They experiment using an $\varepsilon$-greedy algorithm with probability dropping down to the minimum of $0.03$ and not depending on the Q-values, unlike the logit algorithm they claim their explanation relies on, so the latter is not applicable to the profits above the Nash equilibrium they obtained (they are comparable to their memoryless case profits). We do not explain those profits. We did not test a logit exploration rule.

\textbf{Agent-based models.} \citet{tesfatsion2002} describes agent-based computational economics as the study of economies as evolving systems of autonomous agents that interact. In these systems the outcomes come from the interactions of the agents and not from coordination imposed from outside. She asks that such studies test clearly stated hypotheses with controlled and replicable experiments, saved random-number seeds, and openly shared code. The paired worlds with common random numbers in our study, and the replication package, follow this practice. We cite two Iowa State working paper versions (15 March 2002, 31 pages, and revised July 2002, 30 pages), not the published \textit{Artificial Life} article.

\citet{windrum2007} review how agent-based models are validated against data. They recommend sensitivity analysis over the parameters, the initial conditions, and the run-to-run randomness, and also robustness to the noise distribution, the timing of updates, and the level of aggregation. They also report the warning of Cowan and Foray that the counterfactual outputs of agent-based models can mislead as policy counterfactuals, because the systems are stochastic and non-ergodic. In our study the counterfactuals are the known truth inside the model, and they are not claims about a real market. The market is also not validated against data; the robustness checks here are the exploratory runs and the calibration appendix. \citet{fagiolo2007} describe validation as comparing the output of the data generating process of the model with real-world observations, and they list the timing of the model period and the choice of the simulated sub-sample as open problems. They also say that comparing outputs across parameter settings (comparative dynamics) is informative about the model but unclear as a policy prediction, and they report the critique of Cowan and Foray about counterfactuals. We did not read Cowan and Foray, so both accounts of their critique are second-hand. Our comparisons are of the second kind, that is, comparisons inside the model.

\section{Setup}
\subsection{Market and learners}
In our study, $N$ firms choose quantities in a repeated Cournot market with the inverse demand $P=a-bQ$, where $a=100$ and $b=1$, and the marginal cost is $c_i=20+u_i$ with $u_i\sim U[-3,3]$. A treated firm gets a per-unit subsidy $s=5$. The quantities are chosen from a grid of $K=21$ points between zero and $2(a-\bar c+s)/(b(N+1))$. Each firm is a stateless Q-learner with the discount $\gamma=0$, the exploration $\varepsilon_t=\max(\varepsilon_{\min},e^{-\beta t})$ with $\beta=0.003$ and $\varepsilon_{\min}=0.03$, and the learning rate $\alpha=0.05$ (slow) or $0.10$ (fast). A market runs for 20{,}000 periods. The first $T_0=2500$ periods are a burn-in without the policy, the policy is switched on at $T_0$, and the outcome of a firm is its mean profit over the last 500 periods. We chose these settings in a calibration which required the firms to respond to the policy (Appendix~\ref{app:calib}). Since the learners are stateless and $\gamma=0$, the firms do not plan over time.

\subsection{Assignment}
The treatment is assigned at $T_0$ and only the information available at that time is used: the cost type and the burn-in profit, plus noise. The selection strength $\kappa$ controls how strongly the assignment depends on these; $\kappa=0$ is random assignment and $\kappa=1.5$ is strong selection. A fixed number of firms is treated in each market. In the \emph{fixed-share} design half of the firms are treated in every market, and in the \emph{varying-share} design the share is drawn for each market from $\{0.25,0.5,0.75\}$.

\subsection{Effects and ground truth}
Let $D_i$ be the treatment of firm $i$ and $G_i$ the share of the treated rivals, and let $Y_i(d,g)$ be the outcome. The four effects are as follows. The \emph{direct effect} DE is $E[Y_i(1,g)-Y_i(0,g)]$, with the rivals held fixed. The \emph{total effect on treated firms} TE is $E[Y_i(1,G_i)-Y_i(0,0)\mid D_i=1]$. The \emph{spillover} SP is $E[Y_i(0,G_i)-Y_i(0,0)\mid D_i=0]$. The \emph{global effect} GTE is $E[Y_i(1,1)-Y_i(0,0)]$. We use the name DE for the effect that \citet{mkw2025} call the average direct effect, and GTE for their total effect; SP is similar to their indirect effect, but it is not the same.

For every market the simulator runs four worlds from the same burn-in state with the same random draws: no policy, the realized assignment, all firms treated, and the realized assignment with the treatment of one random firm flipped (for DE). The true effects are the averages over the markets. The Monte Carlo errors are below 2 percent of each true value in every cell. With a zero subsidy all the effects are exactly zero, and the Nash benchmark profit is reproduced, as the unit tests check.

\section{Estimators and design}
\textbf{Estimators.} The estimators use the post-period outcome and the pre-treatment covariates (cost, burn-in profit), and the standard errors are clustered by market. (1) Naive: the treated outcome minus the untreated outcome. (2) Within: market fixed effects and covariates. (3) Difference-in-differences (post minus pre, with market fixed effects). (4) Double machine learning (partially linear model, random-forest nuisances), with five-fold cross-fitting where all the firms of a market are in the same fold, and a final regression of the residualized outcome on the residualized treatment. (5) Causal forest (the generalized random forest of the Python package econml, version 0.17.1), with the same folds, using the debiased score for the average effect; its honesty splitting is done by firm and not by market, which is a limitation of the implementation. (6) Exposure-aware versions of (4) and (5), which add the share of the treated rivals $G_i$ as a second treatment; the global effect is the own-treatment coefficient plus the exposure coefficient. These are undefined when the number of treated firms is the same in every market. (7) An OLS regression on the treatment and the exposure without market fixed effects, as an additional linear benchmark.

\textbf{Design of the study.} The factors are the market size $N\in\{4,6,8\}$, the fixed or varying share, the selection $\kappa\in\{0,1.5\}$, and the placebo (policy off, which checks that the estimators do not find effects that are not there). Each of the 24 cells has 50 repetitions of 300 markets, with the learning rate 0.05. We fixed the factor levels, the smallest effect of interest (10 percent of the true effect), the hypotheses, and the decision rules in a dated protocol before a pilot, and we made one amendment after the pilot and before the main run (Appendix~\ref{app:protocol}). The amendment restricted the confirmatory set to slow learners at $N=4,6,8$, because fast learners (learning rate 0.10) and $N\ge 12$ did not meet the convergence criterion of the protocol (at least half of the markets with an unchanged greedy policy). We analyze these cases separately as an exploratory set (Section~\ref{sec:explore}).

\section{Results}
\subsection{The true effects}
Table~\ref{tab:truth} shows the true effects. A subsidy increases the profit of a treated firm and decreases the profit of the untreated firms through a lower price, so SP is negative. The direct effect is larger than the total effect on the treated firms, and the global effect is much smaller than both of them: at $N=4$, treating everyone gives 46 profit units, while treating a single firm and holding the rivals fixed gives 125. The firms earn 1.2 to 1.8 times the Nash profit, so these learners do not behave competitively.
\begin{table}[t]\centering\small\caption{True effects and learning diagnostics (slow learners, fixed share, $\kappa=0$; mean over 50 repetitions of 300 markets).}\label{tab:truth}
\begin{tabular}{rrrrrrrr}\toprule $N$ & DE & TE & SP & GTE & TE$-$SP & Converged & Profit/Nash\\\midrule
4 & 124.6 & 101.5 & -50.0 & 45.6 & 151.5 & 0.76 & 1.22\\
6 & 103.4 & 75.4 & -44.0 & 28.6 & 119.5 & 0.75 & 1.50\\
8 & 88.4 & 59.8 & -39.7 & 19.2 & 99.5 & 0.67 & 1.76\\
\bottomrule\end{tabular}\end{table}
\begin{table}[t]\centering\small\caption{Relative bias (in percent) of the own-treatment coefficient against DE and against TE$-$SP. Slow learners, 50 repetitions per cell. Entries within 10\% in absolute value are the smallest-effect-of-interest cases.}\label{tab:bias}
\begin{tabular}{llrrrrrr}\toprule & & \multicolumn{3}{c}{vs DE} & \multicolumn{3}{c}{vs TE$-$SP}\\\cmidrule(lr){3-5}\cmidrule(lr){6-8} Design, $\kappa$ & Estimator & $N=4$ & $N=6$ & $N=8$ & $N=4$ & $N=6$ & $N=8$\\\midrule
Fixed, 0.0 & Naive & 20 & 16 & 13 & -1 & 0 & 0\\
 & Within & 21 & 16 & 12 & -0 & 0 & -0\\
 & DiD & 22 & 15 & 12 & 0 & -0 & -0\\
 & DML & 21 & 16 & 12 & -1 & 0 & -0\\
 & Forest & 21 & 16 & 12 & -1 & 0 & -0\\
\midrule
Fixed, 1.5 & Naive & 85 & 77 & 70 & 50 & 50 & 50\\
 & Within & 21 & 17 & 15 & -2 & -0 & 2\\
 & DiD & 2 & 4 & 3 & -17 & -11 & -9\\
 & DML & 23 & 18 & 15 & -1 & 0 & 2\\
 & Forest & 22 & 18 & 15 & -1 & 0 & 2\\
\midrule
Varying, 0.0 & Naive & 7 & 9 & -3 & -11 & -6 & -14\\
 & Within & 21 & 16 & 12 & 1 & -0 & 0\\
 & DiD & 21 & 16 & 12 & 0 & -0 & 0\\
 & DML & 6 & 8 & -5 & -12 & -6 & -15\\
 & Forest & 6 & 8 & -5 & -12 & -6 & -15\\
\midrule
Varying, 1.5 & Naive & 60 & 65 & 42 & 30 & 42 & 26\\
 & Within & 20 & 16 & 13 & -2 & -0 & 0\\
 & DiD & -0 & 3 & -1 & -19 & -11 & -12\\
 & DML & -4 & 1 & -18 & -22 & -13 & -27\\
 & Forest & -4 & 1 & -17 & -22 & -13 & -26\\
\bottomrule\end{tabular}\end{table}
\begin{table}[t]\centering\small\caption{Global effect from exposure-aware estimators (designs with varying treated share, slow learners). Relative bias in percent and coverage of the nominal 95\% interval.}\label{tab:gte}
\begin{tabular}{llrrrrrr}\toprule & & \multicolumn{3}{c}{Relative bias (\%)} & \multicolumn{3}{c}{Coverage}\\\cmidrule(lr){3-5}\cmidrule(lr){6-8} $\kappa$ & Estimator & $N=4$ & $N=6$ & $N=8$ & $N=4$ & $N=6$ & $N=8$\\\midrule
0.0 & DML, exposure-aware & -0 & -2 & 9 & 0.94 & 0.88 & 0.88\\
 & Forest, exposure-aware & -0 & -3 & 7 & 0.94 & 0.90 & 0.82\\
 & OLS, exposure & -3 & -16 & -15 & 0.94 & 0.98 & 0.92\\
\midrule
1.5 & DML, exposure-aware & -4 & -16 & -24 & 0.94 & 0.92 & 0.82\\
 & Forest, exposure-aware & -3 & -16 & -25 & 0.94 & 0.92 & 0.82\\
 & OLS, exposure & -6 & -21 & -36 & 0.96 & 0.94 & 0.82\\
\bottomrule\end{tabular}\end{table}

\subsection{What the estimators recover}
\begin{figure}[t]\centering\includegraphics[width=0.95\textwidth]{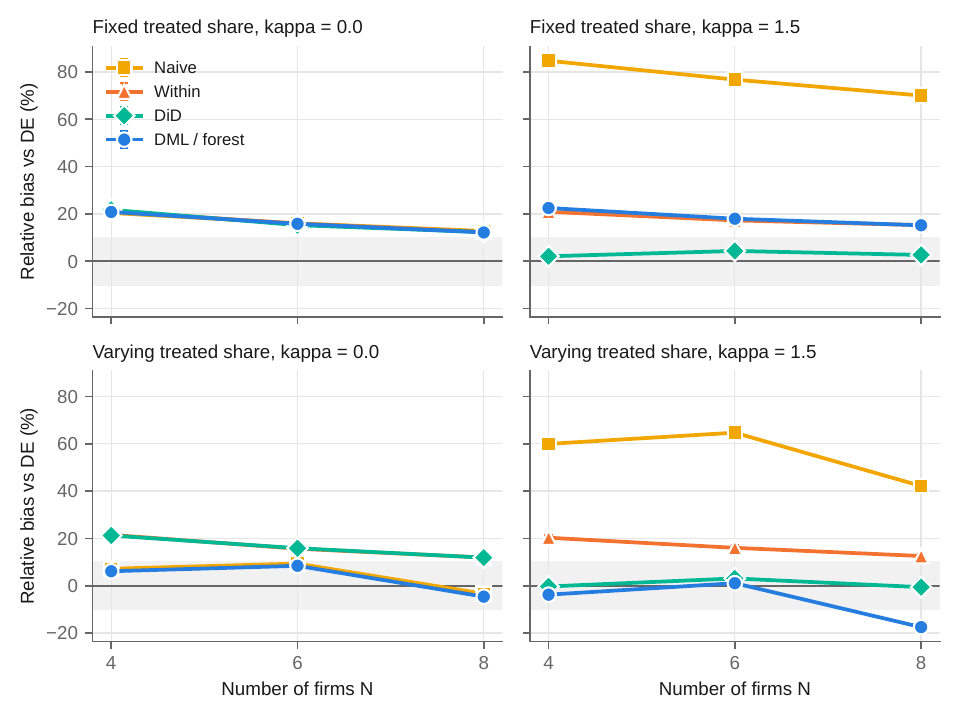}
\caption{Relative bias of the own-treatment coefficient against the direct effect. Shaded band: 10 percent of the true effect. Bars: 95 percent Monte Carlo intervals over 50 repetitions. Within, DiD, DML and the forest coincide in the top-left panel.}\label{fig:bias}\end{figure}

Table~\ref{tab:bias} and Fig.~\ref{fig:bias} show the error of the own-treatment coefficient against DE and against TE$-$SP.
Three results are important.

\emph{First}, when there is no selection into treatment, the estimators that use only the variation within markets (within, DiD, DML and forest) recover TE$-$SP, not DE. With a fixed treated share and no selection, all five estimators are within 1 percent of TE$-$SP and 12 to 22 percent above DE. This is the estimand that the literature on interference leads us to expect for difference-in-differences (\citealp[Proposition 2.1]{butts2023}, where the total effect on the treated and the spillover on the controls correspond to TE and SP defined in Section 3.3). It is also the same kind of target as the difference between the treated and untreated means discussed by \citet{savje2021} and \citet{hu2022}. We read TE$-$SP as the analogue of that target in this market; the literature does not state this for the price-mediated interference, and we did not prove it. In the language of the protocol, the hypothesis H1 (that the adjusted estimators recover DE) is not supported: they are within 10 percent of DE in only 16 of 48 estimator--cell pairs, and never for the within estimator.

\begin{figure}[t]\centering\includegraphics[width=0.55\textwidth]{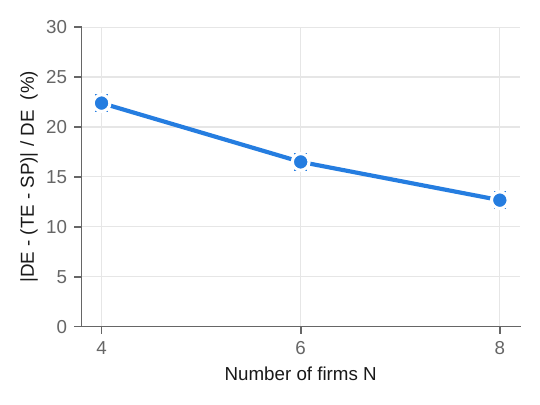}
\caption{Gap between DE and TE$-$SP in percent of DE, by number of firms ($N=4,6,8$; 95 percent intervals over 50 repetitions per cell).}\label{fig:gap}\end{figure}

\emph{Second}, the gap between DE and TE$-$SP is a feature of the finite market and it gets smaller when the market size grows: it is 22.4 percent of DE at $N=4$, 16.5 percent at $N=6$ and 12.7 percent at $N=8$ (Fig.~\ref{fig:gap}; slope $-2.4$ points per firm, clustered by cell, $p<0.001$). This is the direction of the large-market result of \citet{mkw2025}, where the plain contrast gets close to the direct effect. We do not show that the gap disappears, only that it gets smaller from $N=4$ to $8$. One possible reading is that, since a fixed number of firms is treated in each market, a treated firm faces one treated rival fewer than an untreated firm does. In this case the gap shows the spillover of one more treated rival on an untreated firm, and this becomes smaller when the market grows. We have not tested this, and an assignment where each firm is treated independently would be a direct check.

\emph{Third}, selection into treatment ($\kappa=1.5$) biases the naive contrast heavily (42 to 85 percent against DE), while the within estimator is not affected. DML and the forest change little in the fixed-share design (15 to 23 percent against DE), but they go to $-18$ percent against DE at $N=8$ in the varying-share design. Under selection, difference-in-differences is within 5 percent of DE in all six cells ($-0.6$ to $4.3$ percent) and 9 to 19 percent below TE$-$SP, so the statement that it recovers TE$-$SP is true only without selection. Within recovers TE$-$SP under selection in every cell ($-2.3$ to $+2.0$ percent), and so do DML and the forest with a fixed share ($-0.9$ to $+2.1$ percent); with a varying share, DML and the forest are 13 to 27 percent below it. Difference-in-differences also rejects a true null in almost every placebo replication under selection (Section~\ref{app:placebo}). Its accuracy against DE in this case should not be trusted, and we do not know why it fails or why it is near DE.

DML and the forest, which do not use market fixed effects, are within 8 percent of DE in the varying-share designs without selection, where the within estimators stay 12 to 21 percent above it. We do not know if this is because the variation of the treated share across the markets enters these estimates, and we do not claim an explanation.

\subsection{The global effect}
\begin{figure}[t]\centering\includegraphics[width=0.95\textwidth]{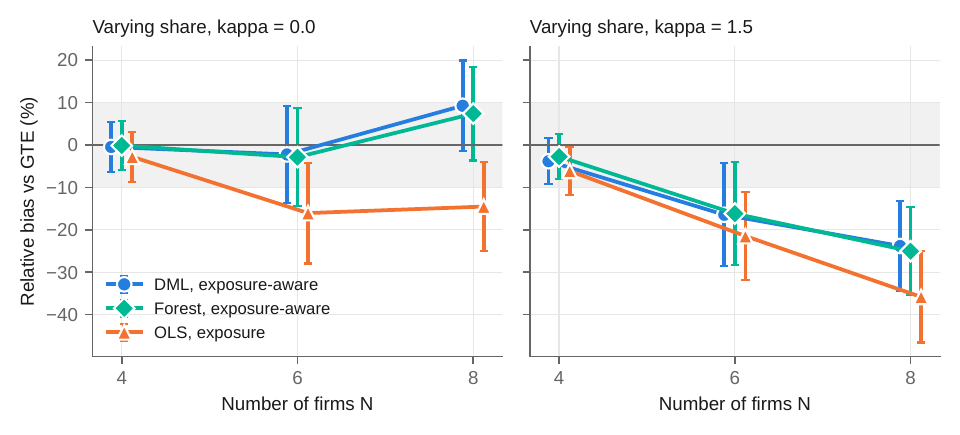}
\caption{Relative bias against the global effect for exposure-aware estimators, designs with varying treated share. Shaded band: 10 percent of the true effect. Bars: 95 percent Monte Carlo intervals over 50 repetitions.}\label{fig:gte}\end{figure}

When the treated share is the same in all markets, the global effect cannot be estimated: the exposure-aware estimators are undefined, and the own-treatment coefficient overstates it by 230 to 430 percent in the fixed-share designs (none of the 30 estimator--cell pairs is within 10 percent, and the same is true for the varying-share designs when the own-treatment coefficient is read as the global effect). When the share varies, the exposure-aware estimators recover it (Table~\ref{tab:gte}, Fig.~\ref{fig:gte}): without selection the relative bias is $-3$ to $+9$ percent for DML and the forest across $N=4,6,8$. With selection the bias grows with the market size, from about $-4$ percent at $N=4$ to $-24$ and $-25$ percent at $N=8$ (DML and forest). Eight of the twelve estimator--cell combinations are within 10 percent (H2; exact binomial $p=0.19$; the Benjamini--Hochberg adjusted value is 0.24, so the data do not let us call this a supported hypothesis). We did not estimate TE with the exposure-aware estimators, so the TE part of H2 is not tested.

\subsection{Learner and design}
The double machine learning and the causal forest give almost the same answer: the mean absolute difference of their own-treatment coefficients is 0.5 profit units, and the mean difference between the fixed-share and varying-share designs is 20 units (H4; bootstrap $p<0.001$). This is true for the average effects only; we do not evaluate heterogeneous effects, where the forests may add more.

\subsection{Coverage and placebo}\label{app:placebo}
The intervals of the own-treatment coefficient reach at least 90 percent coverage of DE in only 1 of 12 cells for DML and 1 of 12 for the forest, and in 0 of 12 for the within estimator. The failure is mostly the bias against the wrong target and not too-narrow intervals: the within estimator covers TE$-$SP in 11 of 12 cells (0.86 to 1.00), and DML covers it at 0.90 to 0.96 in all six fixed-share cells, but only at 0.00 to 0.30 in the six varying-share cells, where its estimate is between DE and TE$-$SP. For the global effect, the exposure-aware intervals cover between 0.82 and 0.94 of the time (Table~\ref{tab:gte}). In the placebo condition, within and DML reject a true null in 0 to 10 percent and 0 to 12 percent of the replications (nominal 5 percent; the Monte Carlo standard error is about 3 points with 50 repetitions). The naive estimator and DiD reject in almost every replication under selection, and the OLS exposure benchmark without fixed effects rejects in every replication with a fixed share, so these should not be used in these settings.

\subsection{Pre-specified hypotheses}
Table~\ref{tab:hyp} summarizes the five hypotheses with the operationalizations in the analysis script.
\begin{table}[t]\centering\small\caption{Hypotheses (confirmatory set: slow learners, $N=4,6,8$). $p$-values are described in the text; BH: Benjamini--Hochberg adjusted.}\label{tab:hyp}
\begin{tabular}{lp{5.4cm}rp{4.6cm}}\toprule
 & Claim (short) & BH $p$ & Verdict\\\midrule
H1 & Adjusted estimators recover DE within 10\%, but not GTE & 0.99 & Not supported (DE: 16/48)\\
H2 & GTE recovered only if the share varies and is used & 0.24 & Not supported at 5\% (8/12 within 10\%)\\
H3 & Coverage below 90\% in some cells & $<0.001$ & Supported (27/36 below 0.90), mostly from bias\\
H4 & DML vs forest gap smaller than the design gap & $<0.001$ & Supported\\
H5 & Gap between DE and TE$-$SP shrinks with $N$ & $<0.001$ & Supported\\\bottomrule
\end{tabular}\end{table}
The operationalizations of H1 and H2 count the estimator--cell pairs as independent units, but they are not independent, so they should be read as summaries of effect sizes and not as exact tests. We wrote the scripts after we had seen the descriptive accuracy tables of the main run but before we computed the tests; H5 was first coded with the wrong sign and we corrected it before reporting.

\subsection{Exploratory set: fast learners and larger markets}\label{sec:explore}
In this part we report the fast learners at $N=4,6,8$ (24 cells, 20 repetitions of 300 markets) and the slow learners at $N=12$ and $N=16$ (20 repetitions of 300 markets per cell, seed offset 7000000). The share of converged markets is important here, because the protocol requires at least 0.50. For the fast learners it is 0.28 to 0.43 per cell (on average 0.43, 0.36 and 0.28 at $N=4,6,8$), and no replication reaches 0.50. For the slow learners it is 0.50 to 0.53 at $N=12$ and 0.36 to 0.39 at $N=16$, and no replication at $N=16$ reaches 0.50. Therefore we treat the fast learners and $N=16$ as a record of what happens in markets that have not converged, not as evidence, and $N=12$ is borderline.

For the fast learners, we used the hypothesis scripts of the main run. Within, DiD, DML and the forest recover DE in 15 of 48 estimator--cell pairs, and all 48 are more than 10 percent from the global effect. Within is within 10 percent of TE$-$SP in all 12 cells, DiD in 6, DML in 7 and the forest in 7. The exposure-aware estimators with a varying share are within 10 percent of the global effect in 5 of 12 cells. The nominal 90 percent intervals cover the direct effect in less than 90 percent of the replications in 29 of 36 cells. The gap between DE and TE$-$SP gets smaller when $N$ grows, from 0.25 to 0.17 to 0.13 of DE. These are counts over estimator--cell pairs, which are not independent.

For the slow learners at $N=12$ and $N=16$, the forest did not run in the first pass, because the package was missing in the environment. We reran $N=12$ with the same seeds, and all the other estimators gave exactly the same results as the first pass. We did not run the forest at $N=16$. Without selection, within, DiD and DML with a common treated share are within $\pm0.4$ percent of TE$-$SP at both sizes and 6 to 9 percent above DE, and the forest at $N=12$ is also close to TE$-$SP ($-0.4$ percent). Therefore the pattern of the main run carries over, with a smaller gap to DE at $N=16$. With selection, DiD is again close to DE ($-12$ to $+4$ percent) and 10 to 12 percent below TE$-$SP, and the naive estimator is 29 to 71 percent above DE. DiD and the naive estimator reject the placebo null in all 20 replications under selection, as in the main run. With a varying share and selection, DML is 32 to 33 percent below DE and the forest at $N=12$ is 33 percent below; we did not examine why. At $N=12$, with a common share and selection, the forest is 21 percent above DE, which is close to DML. The forest rejects the placebo null in 0 percent of the replications without selection and in 15 percent with selection.

The exposure-aware estimates of the global effect do not recover it here. DML with exposure is 23 to 29 percent too small without selection, and the forest with exposure is 33 percent too small at $N=12$. With selection, DML with exposure gives $-6.5$ and the forest gives $-8.0$ at $N=12$, while the true value is 11.2, and at $N=16$ DML with exposure gives $+0.8$ while the true value is 8.4. Since these markets have not converged, we do not read this as a failure of the estimators; it only shows that the finding of the main run was not checked at these sizes or learning speeds.

\section{Discussion}
The main message of the paper is about what is being estimated. A researcher who runs within, DiD, DML or a forest, without selection into treatment, on the data from one market or from markets with a common treated share, gets the total effect on the treated firms minus the spillover. This is the difference between a treated and an untreated firm after the market has reacted. In these markets it is 12 to 22 percent above the direct effect and far above the global effect. If the question is what happens when everyone receives the policy, it is necessary to vary the treated share across the markets and to use an exposure-aware estimator; the estimators themselves (DML or forest) matter much less than this choice of design. The result is more fragile with selection into treatment at larger $N$, where the exposure-aware estimators give a global effect which is too small. In the exploratory run with slow learners at $N=12$ and $N=16$, where the markets had not converged or had only just converged, the exposure-aware estimators did not recover the global effect at all. Therefore the finding for the global effect should not be read beyond $N=8$ and the converged markets of the main run.

\citet{hu2022} note that the direct and indirect effects can depend on the treatment probability of the design and may change sign, and that the data at one treatment rate do not allow a nonparametric estimate of the outcomes at another rate. In our study the global effect is estimated from the shares between $0.25$ and $0.75$, and it is a statement about treating all firms, so it relies on extrapolation by the exposure-aware models. We did not test this with a design that includes the all-treated share. \citet{mkw2025} estimate the indirect effects in one large market by adding small random perturbations to the prices and estimating the price elasticity of the net demand, and they state that this is for markets where no unit has market power and that firms with market power need different techniques. The firms in our study have market power, so we did not apply their design. Whether a perturbation design can be adapted to markets with a few learning firms is an open question.

\section{Limitations}
The results are for one calibration of a stylized market: linear demand, firms which are symmetric up to the cost noise, stateless learners with $\gamma=0$ (no intertemporal planning, no trigger strategies), and the exposure mapping, the share of the treated rivals, which is justified only by the symmetric structure. \citet{leung2020} shows that a response which depends only on the own treatment and the number of treated neighbors follows from local spillovers and exchangeability of peers; here the rivals have different costs, so exchangeability holds only in distribution. The feedback exists because the learners adapt to the subsidy in the post period, and this depends on the learner settings. The firms earn above the Nash profit, so the results describe these learners and not competitive behavior. The hypothesis tests treat the cells as independent and the thresholds are conventions (10 percent of the true effect). The theory of double machine learning which we rely on assumes independent and identically distributed observations (\citealp[Section 3.1]{chernozhukov2018}); the firms in the same market are dependent, so clustering the standard errors and keeping each market in one cross-fitting fold are choices of this study that are not covered by that theory. The same is true for the causal forest: its asymptotic theory assumes independent and identically distributed observations, unconfoundedness and overlap (\citealp[Section 2.1 and Theorem 11]{wager2018}; \citealp[Section 2.1]{athey2019}), and it says nothing about interference. The causal forest has honesty at the firm level. \citet{savje2024} notes that variance estimation under misspecified exposures is not solved and that variance estimators can be anti-conservative, so the coverage of 0.82 to 0.94 which we find for the exposure-aware intervals is a result for these markets and not a property of the estimators. The exposure mapping is not structural in the sense of \citet{leung2026}, and with a fixed number of treated firms the assignments within a market are negatively correlated, so by his example the exposure coefficient does not need to have the sign of the unit-level effects. His results for cluster designs assume Bernoulli assignment within clusters, which is not the case here, and we did not check his other conditions for these markets. The fast learners and the markets above $N=8$ did not reach the convergence criterion, so the confirmatory results do not extend to them. The study has one policy (a per-unit subsidy). We did not evaluate heterogeneous effects, quantile effects, or the estimate of TE by the exposure-aware estimators.

\section{Conclusion}
In a Cournot market with Q-learning firms, where the true effects are known from paired worlds, the estimators we tested recover a quantity which is different from the one a researcher usually wants. Without selection, within, DiD, DML and a causal forest estimate the total effect on the treated firms minus the spillover, and not the direct effect; the gap to the direct effect is 12 to 22 percent in markets of 4 to 8 firms and it gets smaller when $N$ grows. The global effect cannot be estimated when every market has the same treated share. Varying the share across the markets and using an exposure-aware estimator gets close to it in the converged markets of the main run, and this does not carry over to the exploratory markets at $N=12$ and $N=16$. The choice of the design and of the estimand matters more than the choice between DML and a forest. These results are for one stylized market and one calibration, and they do not say how the estimators behave in real data.

\appendix
\section{Calibration of the learners}\label{app:calib}
We chose the learner setting ($\gamma=0$, exploration floor 0.03, 17,500 post-policy periods) after a one-seed grid at $N=6$ with three treated firms, homogeneous costs and 300 markets, over three post-policy lengths, three floors, four learning rates and two discount factors (72 cells, \texttt{scripts/calibrate\_grid.py}). Table~\ref{tab:calib} shows the $\gamma=0$ cells with the chosen floor and length. Lower learning rates converge more often and leave little spillover; faster rates give a large spillover but converge in few markets. The setting is a compromise between the two, and we did not search it more finely. The grid did not use heterogeneous costs; we read it only as a diagnostic.
\begin{table}[h]\centering\small
\caption{Calibration grid, $N=6$, $\gamma=0$, floor 0.03, 17,500 post-policy periods, one seed, 300 markets.}\label{tab:calib}
\begin{tabular}{rrrrr}\hline
$\alpha$ & converged share & TE & SP & GTE \\\hline
0.02 & 0.98 & 55.4 & $-2.7$ & 50.7 \\
0.05 & 0.79 & 72.7 & $-44.7$ & 27.5 \\
0.10 & 0.42 & 82.1 & $-50.9$ & 28.8 \\
0.20 & 0.03 & 85.7 & $-49.3$ & 26.4 \\\hline
\end{tabular}\end{table}
Table~\ref{tab:calibcheck} checks the chosen setting at $N=4,8,16$ with heterogeneous costs, one seed (\texttt{scripts/check\_calibrated.py}). The converged share falls with $N$ and is below 0.50 for fast learners at $N=8$ and $N=16$ and for slow learners at $N=16$ in this check, which is consistent with the converged shares in Section~\ref{sec:explore}.
\begin{table}[h]\centering\small
\caption{Check of the chosen setting, heterogeneous costs, one seed, 300 markets. Profit is relative to the Nash profit.}\label{tab:calibcheck}
\begin{tabular}{rlrrrrrr}\hline
$N$ & learners & converged & profit/Nash & DE & TE & SP & GTE \\\hline
4 & slow & 0.76 & 1.23 & 120.2 & 98.7 & $-49.7$ & 45.3 \\
4 & fast & 0.39 & 1.30 & 144.5 & 113.0 & $-59.3$ & 42.5 \\
8 & slow & 0.68 & 1.75 & 92.3 & 61.6 & $-36.6$ & 18.6 \\
8 & fast & 0.26 & 1.71 & 93.7 & 65.6 & $-42.2$ & 16.9 \\
16 & slow & 0.36 & 2.45 & 54.5 & 22.1 & $-35.0$ & 8.4 \\
16 & fast & 0.06 & 2.15 & 52.1 & 28.0 & $-27.5$ & 9.6 \\\hline
\end{tabular}\end{table}

\section{Protocol and amendment}\label{app:protocol}
The protocol (PROTOCOL.md in the repository at \url{https://github.com/berildemiroglu/cournot-causal-benchmark}, commit cc25fa0 of 27 November 2026; the commit history is a timestamp kept by us, not an independent registry) fixed the factors, the estimators, the smallest effect of interest, the decision rules and hypotheses H1--H5 before the pilot. Addendum 1 (27 November 2026, after the pilot and before the main run) restricted the confirmatory set to slow learners at $N=4,6,8$, because fast learners ($0.28$--$0.44$ converged) and $N=16$ ($0.37$--$0.39$) failed the convergence criterion of at least $0.50$, and $N=12$ was borderline ($0.47$--$0.52$). Thresholds and hypotheses were not changed.

\section*{Statements and Declarations}
\textbf{Funding.} This research did not receive funding.

\textbf{Competing interests.} The author has no competing interests to declare.

\textbf{Data availability.} The simulation code, the analysis scripts, the protocol (PROTOCOL.md, commit cc25fa0), and the raw results of the main run and the exploratory runs reported in this paper are openly available in the GitHub repository \url{https://github.com/berildemiroglu/cournot-causal-benchmark} (release v1.0, commit 58f1ecc, \url{https://github.com/berildemiroglu/cournot-causal-benchmark/releases/tag/v1.0}) under the MIT license. All data are simulated. No human participants or external datasets were used.

\textbf{Author contributions.} Beril Demiroglu is the sole author. She designed the study and its protocol, ran the simulations and the analysis, interpreted the results, and approved the final manuscript. She is accountable for all aspects of the work.

\end{document}